\documentclass[11pt]{article}
\usepackage[utf8]{inputenc}
\usepackage{listings}
\usepackage{xcolor} 
\usepackage{amsmath, amssymb}
\usepackage{graphicx}
\usepackage{hyperref}
\usepackage{geometry}
\usepackage{cite}
\usepackage{float}
\usepackage{algorithm}
\usepackage{algpseudocode}
\title{Development of IoT and AI based Smart Irrigation System}

\author{
    Yunus Emre KUNT\thanks{Email: yunusemrekunt@posta.mu.edu.tr} \\
    Mugla Sitki Kocman University \\
    Mugla, Türkiye\\ 
    \\
    B.Sc. Thesis, Electrical and Electronics Engineering Department \\ \\
    Supervisor: Assist. Prof. Hayriye Serra ALTINOLUK\thanks{Email: serraaltinoluk@posta.mu.edu.tr} 
}
\date{\today}

\begin{document}

\maketitle

\begin{abstract}
Efficient water management is a critical factor in modern agriculture. This study presents a smart drip irrigation system enhanced with Internet of Things (IoT) technologies and artificial intelligence to minimize water waste and improve agricultural productivity. The system is designed to be scalable across various agricultural fields and is prototyped with a test layout. Environmental data are collected via an ESP32 microcontroller and processed using a Long Short-Term Memory (LSTM) model for prediction and autonomous decision-making. Manual and AI-based control are both facilitated through the Blynk application. Key components include rain and soil moisture sensors, a DHT11 temperature and humidity sensor, relays, solenoid valves, and a 12V power supply. The system aims to reduce labor demand in irrigation tasks, support sustainable agricultural practices, and enhance overall crop yield. \\

\end{abstract}

\section{Introduction}
Increasing global population growth, industrialization, and climate change exert significant pressure on freshwater resources, with agriculture accounting for approximately 75\% of global water consumption.\cite{awallace2000} This highlights the urgent need for sustainable water management in the agricultural sector to mitigate the growing water crisis. This issue is illustrated in Figure~\ref{fig:water_demand} below. 
\begin{figure}[H]
\centering
\includegraphics[width=0.85\textwidth]{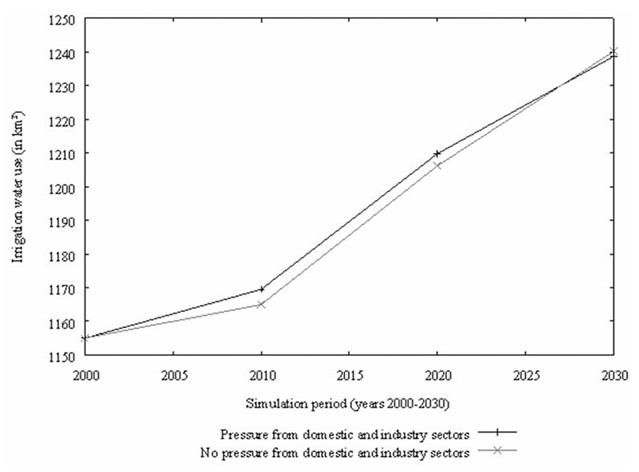}
\caption{Global agricultural water demand projection from 2000 to 2030. Source: \cite{bsauer2010irrigation}.}
\label{fig:water_demand}
\end{figure} 

Particularly in semi-arid regions, efficient irrigation practices are crucial to conserve water while maintaining crop productivity\cite{bsauer2010irrigation},\cite{wada2010global}.

In Turkey, surface irrigation dominates, causing substantial water loss due to evaporation, and shifting toward pressurized systems like drip irrigation could reduce consumption by nearly 50\% \cite{yildirim2012turkiye}. Given Turkey's diverse agro-climatic conditions and crop varieties, improving irrigation efficiency is vital. Drip irrigation, although costly, provides the highest water use efficiency (WUE) compared to other methods. However, it lacks optimal electronic control systems \cite{ertek2014agricultural}.

This study aims to enhance drip irrigation effectiveness by integrating Internet of Things (IoT) and artificial intelligence (AI) technologies, building on historical irrigation developments and addressing contemporary water management challenges.

\section{Methodology}
\subsection{System Architecture}

This study aims to develop a flexible, autonomous, and data-driven irrigation system by integrating Artificial Intelligence (AI) and Internet of Things (IoT) technologies. The primary objective is to enhance irrigation precision, reduce water consumption, and support sustainable agriculture.

The methodology consists of five key stages:

1- Hardware Design and Installation: An ESP32-based circuit was designed, integrating soil moisture, temperature (DHT11), rainfall, and water flow sensors to collect essential environmental data.

2- Data Transmission: Sensor data were transmitted in real time via ESP32 to the Blynk platform for monitoring and were simultaneously stored in Google Sheets for long-term analysis.

3- AI Model Development: An LSTM (Long Short-Term Memory) model was trained using historical sensor data to predict irrigation needs, enabling more efficient and informed watering decisions.

4- Mobile Application Integration: A user-friendly mobile application was developed, supporting manual, automatic (sensor-based), and AI-driven irrigation modes, allowing flexible system control and real-time data access.

5- System Testing and Validation: The entire system was rigorously tested under various conditions to ensure stability, accuracy, and responsiveness. The system also includes fault detection mechanisms to prevent over-irrigation or hardware failures.

\begin{figure}[h]
\centering
\includegraphics[width=0.6\textwidth]{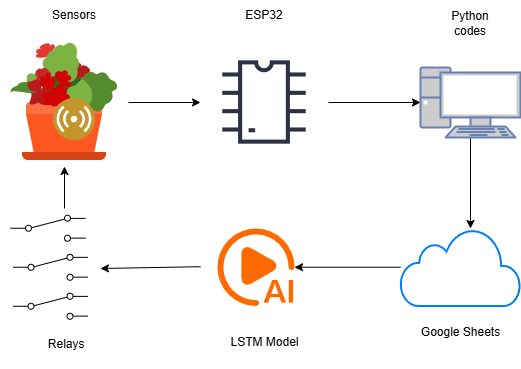}
\caption{General scheme of the proposed system}
\label{fig:drawio_sheme}
\end{figure}

The complete system scheme is summarized in Figure~\ref{fig:drawio_sheme}.

This integrated approach not only promotes water efficiency—especially in water-scarce regions—but also reduces farmers’ workloads and supports sustainable agricultural practices.\\
\subsection{Materials}
The materials and techniques used to develop and deploy an intelligent, self-sufficient irrigation system utilizing AI and the IoT are thoroughly covered in this chapter. While techniques are created to optimize system performance and accuracy, materials are carefully chosen to guarantee compatibility, efficiency, and dependability.\\

1. ESP32:
ESP32 is a low-cost, high-performance microcontroller module developed by Espressif Systems, widely used in IoT and embedded system projects. The dual-core Xtensa LX6 processor offers a flexible and powerful solution for applications requiring wireless communication with integrated Wi-Fi and Bluetooth features\cite{maier2017comparative}\cite{padmasree2023developing}.\\

2. Rain Sensor:
The rain sensor allows the system to adjust irrigation programmes by providing instant updates according to current weather conditions. It stops irrigation when it rains, prevents plants from getting unnecessary water and warns the user.\\

3. Soil Moisture Sensor:
Moisture is an important factor in plant growth. Soil moisture sensors measure the moisture content of the soil and inform users when to start or stop irrigation. In this way, the timing of irrigation is optimized, ensuring efficient use of water resources. Moisture has a critical role in agricultural production as one of the three main factors affecting plant growth\cite{numbi2024development}.\\

4. DHT11 Sensor:
The DHT11 sensor provides a critical layer of environmental data for the irrigation system by measuring ambient temperature and humidity. This sensor enriches the overall dataset, providing important inputs to the AI model to optimise water use\cite{parepalli2024iot}\cite{srivastava2018measurement}.\\

5. YF-S201 Water Flow Sensor:
The YF-S201 water flow sensor is a sensor used to determine the amount of water transported through the pipe by measuring the flow rate of water. This sensor is connected to the pipe of a submersible water pump and has an operating flow rate from 1 to 30 L/min. Furthermore, the water pressure capacity of this sensor is below 1.95 MPa. The YF-S201 improves efficiency by enabling precise management of water in irrigation systems\cite{jones2004irrigation}. \\

6. Solenoid Valve:
Solenoid valves are planned to be operated with 12V DC power supply. By sending a 5V signal to the relay modules, the ESP32 microcontroller will be able to control the valves. This method offers a dependable control mechanism and minimal power consumption, which improves the system's sustainability and efficiency.\\

\begin{figure}[h]
\centering
\includegraphics[width=0.6\textwidth]{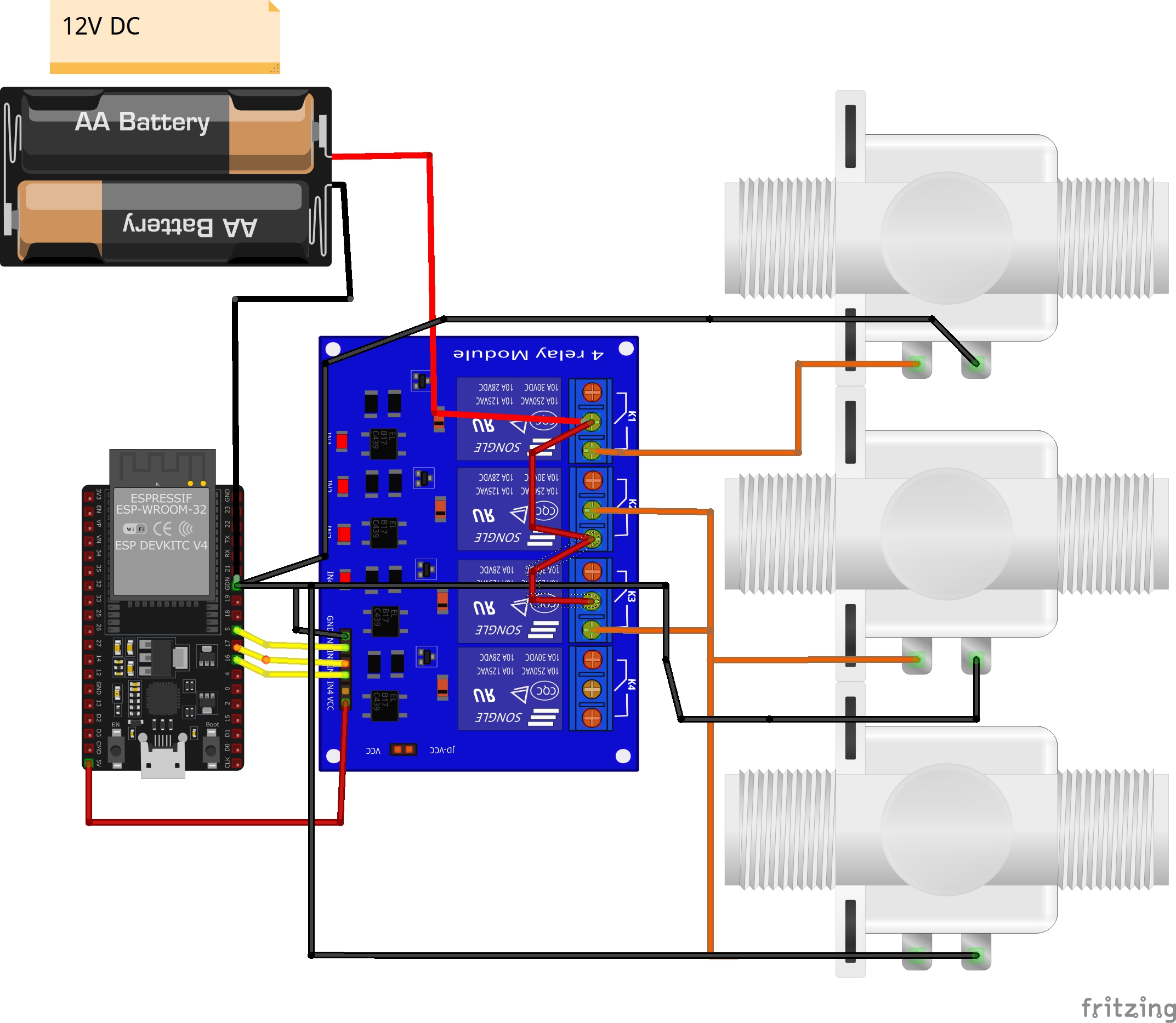}
\caption{Circuit diagram and connection layout of the relays}
\label{fig:relay_connections}
\end{figure}

7. Four-Channeled Relay Module: 
Since each valve requires independent control, a 4-channel relay module is necessary. In manual mode, the valves are controlled via the mobile application; in AI mode, control is based on an LSTM algorithm; and in auto mode, operation depends on predefined threshold values. Figure ~\ref{fig:relay_connections} illustrates the circuit diagram and connection layout of the relays which control solenoid valves.

\subsection{Methods}
\begin{figure}[h]
\centering
\includegraphics[width=0.6\textwidth]{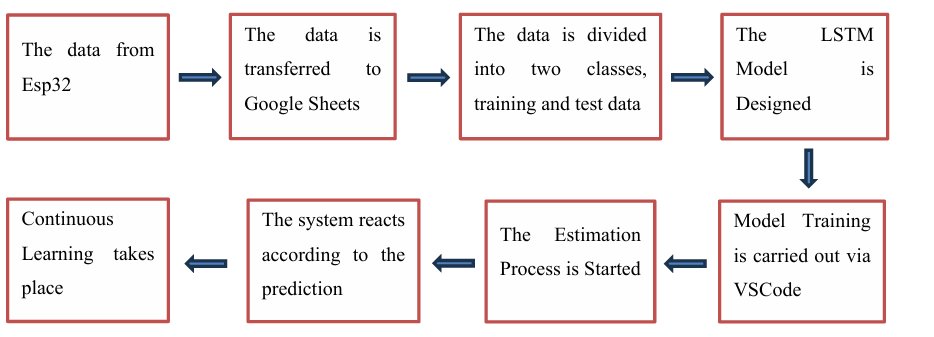}
\caption{Scheme of the data management process}
\label{fig:work}
\end{figure}

\begin{figure}[h]
\centering
\includegraphics[width=0.6\textwidth]{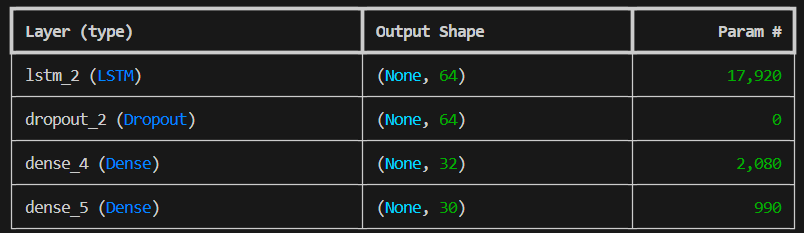}
\caption{ The architecture of the ML model}
\label{fig:table}
\end{figure}

\begin{figure}[h]
\centering
\includegraphics[width=0.6\textwidth]{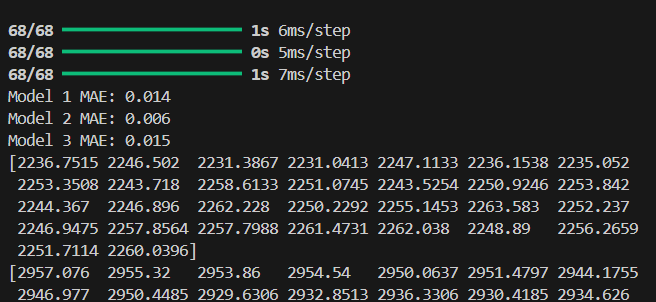}
\caption{The corresponding MAE values}
\label{fig:mae}
\end{figure}

\begin{figure}[H]
\centering
\includegraphics[width=0.6\textwidth]{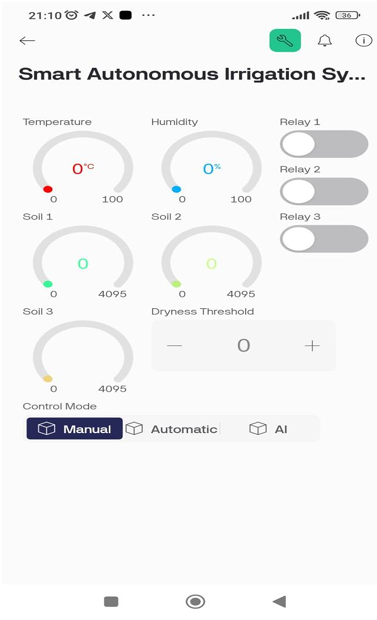}
\caption{ Interface of mobile application}
\label{fig:blynk}
\end{figure}

\textbf{Step 1: System Design and Circuit Board Construction}\\
The first stage of the project, a circuit board was designed to control the irrigation and 
data collection processes. Several sensors are included on this board to monitor 
environmental conditions and manage the irrigation system.\\
\textbf{Step 2: Data Transmission and Cloud Storage} \\
The ESP32 microcontroller was programmed via Arduino IDE to collect sensor data. 
This data was then read through a Python script over a serial connection. Using the 
ESP32’s wireless capabilities, the processed sensor data was transmitted in real time 
to cloud-based platforms, including Blynk for monitoring and Google Sheets for data 
logging and storage. \\
\textbf{Step 3: System Control Modes} \\
The system will offer three control modes to manage the irrigation process effectively: 
\\1. Manual Mode: Users have full control over the irrigation system and can start 
or stop the irrigation process through a mobile application. This mode provides 
complete flexibility for users to manage the irrigation manually. \\
 2. Automatic Mode: In automatic mode, the system autonomously initiates or 
terminates irrigation based on sensor data. In the initial phase, a fixed threshold 
value. When the sensor value exceeds this threshold (indicating dry soil), the system 
starts irrigation; when the value falls below the threshold (indicating sufficient 
moisture or rainfall), irrigation is stopped.\\
3. AI Mode: In this mode, a machine learning algorithm, specifically an LSTM 
model, was implemented to predict irrigation needs based on historical data. 
\textbf{Environmental data} (such as temperature, humidity, and soil moisture), along 
with user-specific water consumption records, were collected and stored during 
the project. This dataset was used to train the \textbf{LSTM model}, enabling the 
system to generate personalized irrigation predictions. The AI mode 
\textbf{successfully analyzed} past irrigation patterns and environmental conditions, 
providing users with a more efficient and adaptive watering schedule. \\
\textbf{Step 4: Irrigation Mechanism}\\ 
In the completed system, the ESP32 microcontroller managed the irrigation process 
by controlling solenoid valves through relay modules. These solenoid valves, powered 
by a stable 12V DC power supply, successfully distributed water to designated zones 
based on real-time soil moisture data. This structure enabled targeted and efficient 
irrigation for each area. \\
A constant 5V DC power supply was used to ensure the stable operation of the ESP32 
microcontroller, sensors, and relay modules. With uninterrupted power and automated 
valve control, the system operated continuously without the need for manual 
intervention, providing a reliable and autonomous irrigation solution. \\
\textbf{Step 5: System Monitoring and Data Storage}\\
 Sensors data transferred to the Google Sheet by using a Python script. These data are subsequently utilized for model training and serve as a critical component in enabling the model's \textbf{continuous learning} and \textbf{adaptive capabilities}. The data is split into three sets: training, validation, and test. 70\% of the data 
is allocated to the training set to train the model, 15\% to the validation set, which is 
used during training to fine-tune the model and differs from the test set, and the 
remaining 15\% to the test set, which is used to evaluate the system’s accuracy. 
For ongoing monitoring and analysis, all sensor data and system control activities was 
saved in the cloud. Users are able to assess the system's performance over time and 
make the required modifications to maximize its effectiveness thanks to the saved data. 
Based on current circumstances, this data is also being utilized for system performance 
improvement and troubleshooting. Figure~\ref{fig:work} illustrates the data management process as a scheme\\ 
\\
\textbf{Forecasting Process with ML Models} \\
In this project, LSTM algorithm, one of the machine learning methods, was used to 
estimate the irrigation requirement. LSTM is an enhanced variant of recurrent neural networks (RNNs), specifically designed to capture \textbf{complex dependencies} and long-term temporal relationships in time series data. While traditional RNNs face problems such as gradient fading and gradient 
bursting in time-delayed data sequences, LSTM has a unique cell structure to solve 
these problems. This cell structure is equipped with input, forget and output gates 
that control the flow of information and thus effectively learns both short and long
term dependencies in data sequences\cite{sherstinsky2020fundamentals}. \\\\\\
\\
\textbf{LSTM Model Design and Training}\\
The pre-processed data is uploaded to VSCode for the design and training of the 
LSTM model using Python. VSCode provides a powerful environment for machine 
learning tasks and offers a user-friendly interface that facilitates easy code and library 
management. During model development, the most commonly used library was 
TensorFlow due to its capabilities for building LSTM models with layered structures. 
TensorFlow also includes the Keras API, which simplifies the model development 
process significantly. The LSTM model is designed to predict irrigation requirements 
based on environmental data. \\
In the training phase, \textbf{hyperparameters} such as features, targets, and layers are tuned 
with various values to identify the optimal model architecture. In machine learning, 
'features' refer to the input variables that influence the model training, while 'targets' 
are the output variables the model aims to predict. For example, when identifying a 
glass, the presence of a handle could be a feature, and predicting whether it is for hot 
or cold drinks could be a target. In this model, features include temperature, humidity, 
and soil moisture sensor data, while targets correspond to future soil moisture values. 
Three separate models correspond to three individual pots, each with distinct feature 
values. Therefore, the same LSTM architecture is applied under three different model 
names to accommodate these variations. The Rectified Linear Unit (ReLU) 
activation function is employed to improve \textbf{model accuracy}.  \\
\textbf{ReLU} is a widely used activation function in deep learning that accelerates training 
and enhances performance by converting negative values to zero, thereby enabling 
faster and more stable learning by effectively managing the model’s learning rate \cite{agarap2018deep}. 
MSE loss function is being used. \textbf{MSE} is a loss function that helps the model to 
minimise the prediction errors. Since this function increases the error values 
quadratically, larger errors are given more weight. Thus, the model performs a more 
precise and careful learning process \cite{wang2021lstm}.\\
Layers also play a critical role in training the model, as the learning process is 
structured around them. Keras offers a variety of layers, including Dense, LSTM, and 
Dropout layers. 
The Dense layer helps the model learn relationships between input features. For 
instance, it can infer that "as temperature increases, soil moisture tends to decrease" 
by identifying correlations within the data. 
The LSTM layer is composed of gates—specifically, the input, forget, and output 
gates—which manage the flow of information through the network. These gates utilize 
activation functions such as \textbf{sigmoid and tanh} to determine which values should be 
retained or discarded, enabling the model to focus on relevant patterns in sequential 
data\cite{zheng2024adaptive}. The \textbf{Dropout layer} is used to prevent overfitting by randomly disabling a portion of 
neurons during training. This encourages the model to generalize better rather than 
memorize the training data, thus improving performance on unseen data\cite{bhimavarapu2023improved}.  \\\\
\textbf{Success Criteria}\\
After the training process is completed, the performance of the model is evaluated on 
test data and the prediction accuracy is measured. The error rate of the prediction is 
evaluated using the MAE metric and this value between 0 and 0.5 is the range 
considered successful by artificial intelligence developers and in this model we have 
‘0.014’. Figure~\ref{fig:table} illustrates the architecture of the ML model, while Figure ~\ref{fig:mae}
presents the corresponding MAE values.
\\
\textbf{Visualisation and Control of Data} \\
The Blynk mobile application, developed within the scope of the project, enables users to easily monitor data from devices such as soil moisture sensors, DHT11 sensors, and rain sensors.
At the same time, irrigation system relays and valves can be controlled directly from Blynk, offering the user maximum flexibility.\\
As the system operates, it continuously collects environmental data—including 
temperature, humidity, rainfall, and soil moisture levels—from various sensors. This 
data is sent from the ESP32 to the backend and subsequently forwarded to the Blynk 
mobile application. Through this interface, users can monitor real-time sensor readings 
and remotely control the system. Figure 24 shows the mobile application interface used 
for monitoring and control.
The design of the Blynk mobile dashboard is presented in detail in Figure ~\ref{fig:blynk}

\section{Results and Discussion}
In this chapter, the outcomes of the simulation, prototyping, and performance 
evaluation processes of the ESP32-based smart irrigation system are presented. The 
system was first tested through virtual simulation using the Wokwi platform to validate 
its core functionalities. Subsequently, a physical prototype was assembled, and the 
irrigation infrastructure was implemented with a focus on robustness and reliability. 
The system’s decision-making capability was assessed through model performance 
analysis using moisture data, while the implementation of multiple operating modes 
and a notification system was also examined to ensure user flexibility and system 
resilience.

\subsection{Physical Prototyping and Wiring  }

\begin{figure}[H]
\centering
\includegraphics[width=0.6\textwidth]{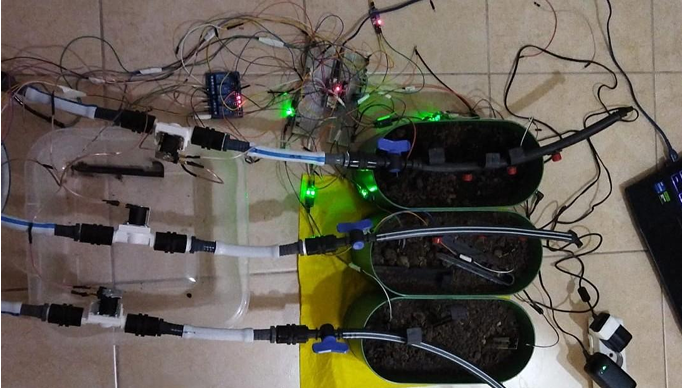}
\caption{ Breadboard wiring of the ESP32-based smart irrigation system with 
relay and solenoid valve connections}
\label{fig:system}
\end{figure}

Following the successful simulation and initial software validation of the smart 
irrigation system, a physical prototype was constructed to verify the hardware design 
and overall functionality. 
The hardware setup consisted of an ESP32 microcontroller interfaced with multiple 
sensors and actuators to control irrigation zones effectively. Three 12V solenoid valves 
were connected via a 4-channel 5V relay module to enable switching of water flow in 
different zones based on sensor inputs. 
The wiring configuration was carefully designed to ensure safe and reliable operation 
as follows: \\
1- The positive terminal (+) of each solenoid valve was connected to the COM 
(common) terminal of the corresponding relay channel via jumper wires.\\ 
2- All relay COM pins were connected together and linked to the positive terminal 
of a 12V external power supply.\\
3- The negative terminal (–) of each solenoid valve was connected to the 
Normally Open (NO) pin of the respective relay. \\
4- All valve negative lines were combined and grounded to the negative terminal 
(GND) of the power supply. \\
5- The ESP32, relay module, and power supply shared a common ground to 
maintain consistent voltage references and avoid floating potentials. \\
This wiring scheme allowed the low-voltage control signals from the ESP32 to safely 
drive the high-current solenoid valves via the relays. The relay module acted as an 
interface, protecting the microcontroller while ensuring reliable valve activation. 
The hardware connections, including the breadboard layout, relay module wiring, and 
solenoid valve integration, are shown in Figure~\ref{fig:system}. This figure illustrates the detailed 
physical setup and the interaction between the microcontroller, relays, and valves.

\subsection{Design and Assembly of the Irrigation System Hardware }
\begin{figure}[H]
\centering
\includegraphics[width=0.6\textwidth]{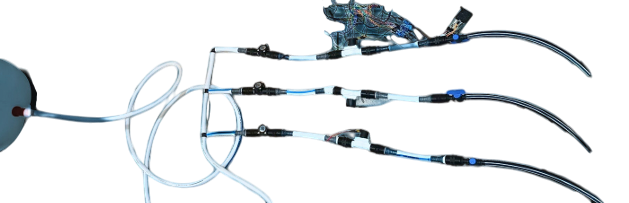}
\caption{General View of the Drip Irrigation System}
\label{fig:harware}
\end{figure}

During the installation of the irrigation system, 16 mm drip fittings and 1/2” (12.5 mm) 
steel reinforced water hose were used in combination to create a leak-proof and durable 
water distribution line. The installation started with the installation of one 16 mm drip 
elbow and two 16 mm drip T-pieces. A 15 cm length of hose was connected between 
the T-pieces, and a 15 cm piece of hose was added between the T-piece and the elbow. 
The 16 mm drip elbow was connected to a 1/2” PVC hose fitting via a 10 cm long 
piece of hose. To ensure tightness, a (½“ x ½”) threaded sleeve was placed between 
the fitting and the system, and Bay-Tec brand Teflon tape measuring 8 m x 12 mm 
was used at the connection points. A water flow sensor was integrated into the 
continuation of this line. After the water flow sensor, the same connection procedures\\
At the end of each row, a Mini Valve Dovetail (16 mm x ½” male) was installed. 16 
mm drip pipes cut into 35 cm lengths were connected to these valves. Three holes were 
drilled on each drip pipe using the Drip Irrigation Red Dripper Nozzle Drill, and three 
drippers were placed in these holes. For each pot, three Drip Irrigation Pipe Fixing 
Stakes were used to fix the drip pipes and ensure that the irrigation reached the right 
points.\\
As shown in Figure~\ref{fig:harware} , the general view of the drip irrigation system.

\subsection{Model Performance Evaluation}
\begin{figure}[H]
\centering
\includegraphics[width=0.6\textwidth]{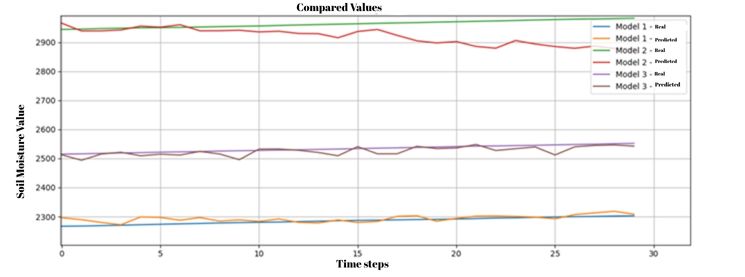}
\caption{ Real and predicted values for each pot}
\label{fig:predicted}
\end{figure}

\begin{figure}[H]
\centering
\includegraphics[width=0.6\textwidth]{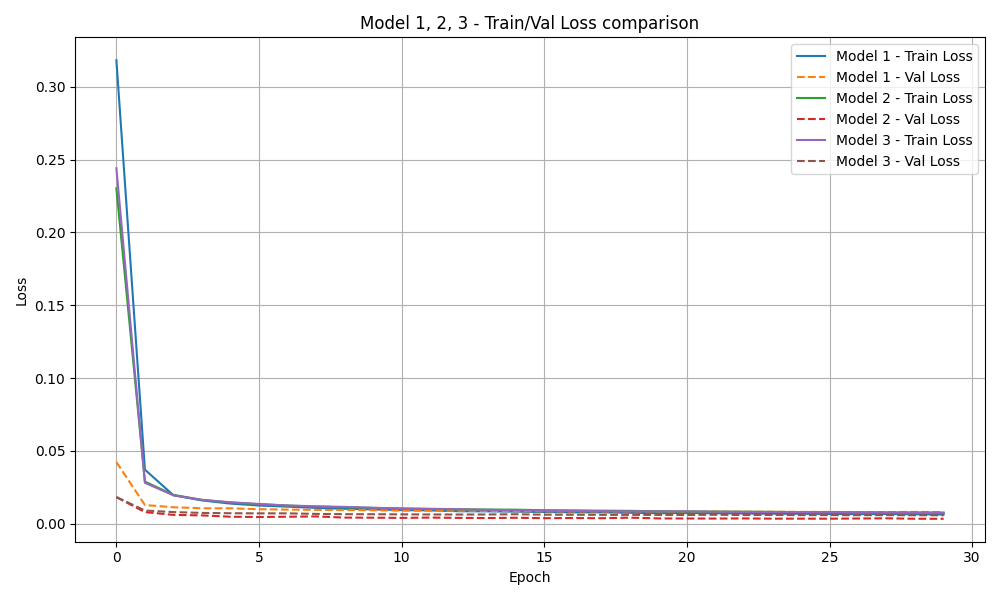}
\caption{ Loss values in a graph}
\label{fig:loss}
\end{figure}

Target list for pot\_1, pot\_2 and pot\_3 contain different moisture levels. The Python 
code decides how long to irrigate by reading the values in the 'target' list, which 
contains data predicted by the model for each pot, according to whether they are below 
or above the specified threshold. Threshold value can be adjusted remotely by Blynk 
application. In Figure~\ref{fig:predicted} , the predicted and actual moisture values for all three trained 
models are compared to evaluate their performance. Figure~\ref{fig:loss} presents the training 
and validation loss values during the training process.

\subsection{Operating Modes and Notification System }
\begin{figure}[H]
\centering
\includegraphics[width=10cm, height=15cm]{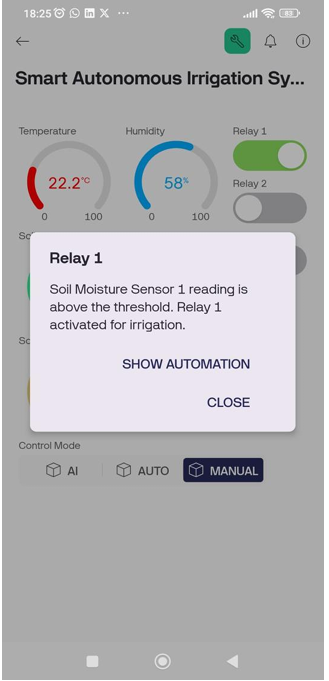}
\caption{  Mobile notification sent via the Blynk platform upon relay activation}
\label{fig:application}
\end{figure}

The related system operates in three different modes: AI, AUTO, and MANUAL. The user can switch between these modes via the mobile application interface. The selected mode is then transmitted to the system, which updates its behavior accordingly. This functionality provides significant flexibility to the user. In case of a malfunction or unexpected behavior, the system can be switched to AUTO or MANUAL modes, ensuring both operational stability and system reliability. This modular control structure allows for seamless transitions between automation levels, offering both intelligent decision-making through AI and manual options when necessary. Even if the Wi-Fi connection is lost, the system continues to operate in the previously selected mode. In case of sensor failure, the system halts irrigation and notifies the user to prevent overwatering. For instance, in AI and AUTO modes, the ESP32 microcontroller executes irrigation decisions locally based on the most recent data or thresholds. However, in MANUAL mode, irrigation halts to prevent unintended operation, as manual control requires user input via the mobile application Application and notification systems are illustrated in the figure~\ref{fig:application} below

\section{Conclusion}
In this thesis, a smart irrigation system integrating LSTM-based soil moisture 
prediction with an ESP32-controlled hardware platform was designed, implemented, 
and validated. The system effectively monitors environmental conditions and 
autonomously controls irrigation valves to optimize water usage.  
Key contributions include the development of accurate per-pot LSTM models that 
allow precise irrigation scheduling, integration of multiple sensors for real-time data 
acquisition, and the creation of a mobile application enabling remote monitoring and 
control with user-adjustable thresholds.  
Experimental results demonstrated reliable system performance, with automatic 
irrigation triggered appropriately according to soil moisture predictions. The 
protective design of the hardware ensured robustness against outdoor environmental 
factors, enabling practical deployment.  
This work provides a scalable and efficient solution for precision agriculture, offering 
potential for large-scale applications in diverse agricultural environments. The 
system’s adaptability and user-friendly interface pave the way for further 
enhancements and integration with advanced AI models.\\

\section*{Acknowledgement}
I would like to extend my deepest gratitude to Ass. Prof. Hayriye Serra Altınoluk 
for her invaluable support, insightful guidance, and encouragement throughout this 
study. Her expertise in IoT-based systems and AI applications, along with her 
constructive feedback, has greatly contributed to the development and successful 
completion of this project. 
I am especially grateful for her dedication and the knowledge she generously shared, 
which helped me overcome challenges and achieve my project goals. 
I also acknowledge the resources and facilities provided by Muğla Sıtkı Koçman 
University, which played a vital role in enabling the successful realization of this study.

\appendix

\clearpage  % Yeni sayfa başlatır
\section{Appendix}

% Renk tanımları
\definecolor{codegray}{rgb}{0.5,0.5,0.5}
\definecolor{codegreen}{rgb}{0,0.6,0}
\definecolor{codeblue}{rgb}{0.2,0.2,0.8}

% Pseudo kod dili tanımı
\lstdefinelanguage{pseudocode}{
  keywords={BEGIN, END, IF, THEN, ELSE, ELSEIF, ELSE IF, WHILE, FOR, FUNCTION, CALL, RETURN, AND, OR, NOT, EQUAL},
  morekeywords={[2]serial,relay,sensor,DHT11,threshold,mode,LOW,HIGH,ON,OFF},
  keywordstyle=\color{codeblue}\bfseries,
  keywordstyle=[2]\color{codegreen},
  comment=[l]{//},
  morecomment=[s]{/*}{*/},
  commentstyle=\color{codegray}\ttfamily,
  basicstyle=\ttfamily\footnotesize,
  sensitive=false,
}

% Kod stili ayarları
\lstset{
  backgroundcolor=\color{gray!10},
  basicstyle=\ttfamily\small,
  keywordstyle=\color{blue},
  stringstyle=\color{red},
  commentstyle=\color{green!60!black},
  numbers=left,
  numberstyle=\tiny\color{gray},
  stepnumber=1,
  breaklines=true,
  frame=single,
  captionpos=t
}

% --- Kod 1: ESP32 C++ Setup ---
\begin{lstlisting}[language=pseudocode, caption={Pseudocode of ESP32 Firmware Logic}, label={lst:dht11}]
BEGIN SETUP
    Initialize serial communication
    Start DHT11 sensor
    Set soil, flow, relay, and rain sensor pins
    Set relays to OFF (HIGH)
END SETUP

BEGIN LOOP
    CALL handleSerialCommands()
    CALL updateRelays()
    IF current time - last send > interval THEN
        Update time
        CALL sendSensorData()
    ENDIF
END LOOP

FUNCTION updateRelays()
    FOR each relay
        IF mode is AUTO THEN
            Read two soil sensors
            Calculate average
            IF average > threshold THEN
                Turn relay ON (LOW)
            ELSE
                Turn relay OFF (HIGH)
            ENDIF
        ENDIF
    ENDFOR
END FUNCTION

FUNCTION handleSerialCommands()
    WHILE serial input available
        Read line
        Update mode, threshold, or relay state accordingly
    ENDWHILE
END FUNCTION

FUNCTION sendSensorData()
    Create JSON object
    Read DHT11, rain, soil, flow sensors
    Append relay and mode info
    Send via serial
END FUNCTION
\end{lstlisting}

% --- Kod 2: LSTM Modeli ---
\begin{lstlisting}[language=pseudocode, caption={Pseudocode of LSTM-Based Moisture Prediction Model}, label={lst:lstm-model}]
BEGIN DATA ACQUISITION
    Authenticate Google Sheets API
    Load sensor data
    Calculate zone-wise average soil moisture
END

BEGIN PREPROCESSING
    Normalize with MinMaxScaler
    Generate sequences: lookback = 60, forecast = 30
    Train/Val/Test split
END

FOR each zone DO
    DEFINE LSTM model:
        Input: 60 timesteps, 5 features
        LSTM + Dropout + Dense + Output
    Compile and train model
    Save best model with checkpoint
END FOR

BEGIN VISUALIZATION
    Plot loss curves
END
\end{lstlisting}

% --- Kod 3: AI Mode Logic ---
\begin{lstlisting}[language=pseudocode, caption={AI Mode Logic for Smart Irrigation System}, label={lst:smart-irrigation-ai}]
BEGIN MAIN
    IF models exist THEN
        Load them
    ELSE
        CALL retrain_models()
    ENDIF

    CALL authenticate_with_google_sheets()
    Fetch and preprocess data
    Normalize with MinMaxScaler
END MAIN

FUNCTION CreateSequenceDataset()
    Generate (X, y) for LSTM
    RETURN X, y
END FUNCTION

BEGIN DATA_SPLIT
    Train, Val, Test split
END DATA_SPLIT

BEGIN PREDICTION
    Predict soil moisture
    Inverse transform results
    Evaluate using MAE
END PREDICTION

FUNCTION RunAIMode()
    FOR each pot
        IF forecast shows need for watering THEN
            Calculate time
            Print decision
        ENDIF
    ENDFOR
END FUNCTION

FUNCTION RunAutoMode()
    FOR each threshold condition
        Turn relay ON then OFF
    ENDFOR
END FUNCTION

FUNCTION RunManualMode()
    FOR each valve
        Read controller input
        Set valve accordingly
    ENDFOR
END FUNCTION
\end{lstlisting}

% --- Kod 4: Blynk Kontrol Sistemi ---
\begin{lstlisting}[language=pseudocode, caption={Blynk-Based Mode Control Logic}, label={lst:blynk-control}]
BEGIN INITIALIZATION
    Import BlynkLib, serial, time
    Import AI/Auto/Manual functions
    Set BLYNK_AUTH and connect
    Initialize serial and mode states
END INITIALIZATION

ON VIRTUAL PIN V9 WRITE
    Update current_mode
END

ON VIRTUAL PINS V5--V7 WRITE
    Update relay states and send serial command
END

BEGIN MAIN LOOP
    CALL blynk.run()
    IF current_mode NOT EQUAL last_mode THEN
        SELECT mode:
            1 --> AI --> run_ai_mode()
            2 --> AUTO --> run_auto_mode()
            3 --> MANUAL --> run_manual_mode()
        Update last_mode
    ENDIF
END MAIN LOOP
\end{lstlisting}
\end{document}